\documentclass{IEEEcsmag}

\usepackage[colorlinks,urlcolor=blue,linkcolor=blue,citecolor=blue]{hyperref}

\usepackage{upmath}

\usepackage{todonotes}
\usepackage{soul}
\usepackage[edges]{forest}
\usepackage[most]{tcolorbox}
\usepackage{tabularx}
\let\labelindent\relax

\usepackage[inline]{enumitem}

\usepackage{tikz}
\usetikzlibrary{arrows,positioning,automata,shapes,arrows.meta,decorations.markings,petri,calc,shadings,patterns,decorations.pathreplacing}

\usepackage{MnSymbol}

\newtcolorbox{glossarybox}[2][]{
                lower separated=false,
                colback=white!80!gray,
colframe=white, fonttitle=\small\bfseries,
colbacktitle=white!50!gray,
coltitle=black,
enhanced,
attach boxed title to top left={xshift=0.4cm,yshift=-2mm},
title=#2,#1}
\newtcolorbox{metricbox}[2][]{
                lower separated=false,
                colback=white,
colframe=black,fonttitle=\bfseries,
colbacktitle=black,
coltitle=white,
enhanced,
attach boxed title to top left={yshift=-0.1in,xshift=0.15in},
                 boxed title style={boxrule=0pt,colframe=white,},
title=#2,#1}

\usepackage[percent]{overpic}

\pgfdeclarelayer{bg}    
\pgfsetlayers{bg,main}  

\usepackage{booktabs}

\usepackage{rotating}
\usepackage{multirow}	
\usepackage{adjustbox}
\usepackage{array}
\usepackage{longtable}
\newcolumntype{R}[2]{%
    >{\adjustbox{angle=#1,lap=\width-(#2)}\bgroup}%
    l%
    <{\egroup}%
}

\usepackage{makecell}

\usepackage{makecell}

\usepackage{cellspace}
\usepackage[printonlyused]{acronym}
\acrodef{cps}[CPS]{cyber-physical system}
\acrodef{dt}[DT]{digital twin}
\acrodef{sedt}[SEDT]{security-enhancing digital twin}
\acrodef{iot}[IoT]{Internet of Things}
\acrodef{ids}[IDS]{intrusion detection system}
\acrodef{ad}[AD]{anomaly detector}
\acrodef{ot}[OT]{Operational Technology}
\acrodef{apt}[APT]{advanced persistent threat}
\acrodef{ics}[ICS]{industrial control system}
\acrodef{sil}[SIL]{software-in-the-loop}
\acrodef{hil}[HIL]{hardware-in-the-loop}
\acrodef{plc}[PLC]{programmable logic controller}
\acrodef{mtd}[MTD]{moving target defense}

\usepackage{pifont}

\jvol{}
\jnum{}
\paper{}
\jmonth{}
\jname{Security \& Privacy}
\pubyear{2023}

\begin{document}

\title{Security-Enhancing Digital Twins: Characteristics, Indicators, and Future Perspectives}

\author{\textsuperscript{\textdagger}Matthias Eckhart (corresponding author), \textsuperscript{$\ddagger$}Andreas Ekelhart, \textsuperscript{$*$}David Allison, \textsuperscript{$\|$}Magnus Almgren, \textsuperscript{$\mathsection$}Katharina Ceesay-Seitz, \textsuperscript{$\mathparagraph$}Helge Janicke, \textsuperscript{$\dagger\dagger$}Simin Nadjm-Tehrani, \textsuperscript{$\ddagger\ddagger$}Awais Rashid, \textsuperscript{$\circ$}Mark Yampolskiy}
\affil{\textsuperscript{\textdagger}SBA Research, \textsuperscript{$\ddagger$}Christian Doppler Laboratory for Security and Quality Improvement in the Production System Lifecycle, University of Vienna, \textsuperscript{$*$}AIT Austrian Institute of Technology, \textsuperscript{$\|$}Chalmers University of Technology, \textsuperscript{$\mathsection$}CERN, \textsuperscript{$\mathparagraph$}Cyber Security Cooperative Research Centre and Edith Cowan University, \textsuperscript{$\dagger\dagger$}Linköping University, \textsuperscript{$\ddagger\ddagger$}University of Bristol, \textsuperscript{$\circ$}Auburn University}

\begin{abstract}
The term \emph{``digital twin''} (DT) has become a key theme of the cyber-physical systems (CPSs) area, while remaining vaguely defined as a virtual replica of an entity.
This article identifies DT characteristics essential for enhancing CPS security and discusses indicators to evaluate them.
\end{abstract}

\maketitle

\chapterinitial{Digital twins} are considered to be a key enabler for future industrial automation technologies.
Driven by Industry 4.0 and ``Factory of the Future'' initiatives, \acp{dt} have evolved from highly specialized aerospace applications to a wide variety of domains, including manufacturing, energy, and transportation.

The interpretation of the \ac{dt} concept varies among researchers~\cite{Negri2017}: some understand it as a digital representation based on a data-driven solution or simulation; others consider it to be a composition of physical models of interdependent components that use input data from the real world to reflect the system's current state or forecast its future behavior.
In the past few years, numerous \ac{dt} applications have appeared in literature, some of which were even further developed and released as a product for market use.
Examples of how \acp{dt} are currently applied in practice include: machine learning models, asset-related data objects (à la asset administration shell), backends for IoT devices (e.g., Eclipse Ditto), virtual testing solutions, and detection algorithms for abnormalities (as presented, for instance, by Jiang et al.~\cite{Jiang2021}).
While recent surveys on the state of \ac{dt} research and technology adoption (e.g.,~\cite{Negri2017, Valk2020, Jones2020}) attempt to consolidate existing definitions and clarify the characteristics of \acp{dt}, they do not consider the use of this concept for security applications.

The capabilities promised by existing implementations of the \ac{dt} concept raise the question of how such virtual replicas of \acp{cps} can also be used for security-enhancing purposes.
The absence of a classification system makes it difficult to recognize the potential of \acp{sedt} and to compare with existing solution proposals, especially if ``digital twin'' is used as an umbrella term for various \ac{cps}-focused security mechanisms.
This gap in the literature hinders progress, as misinterpretation may arise if the proposed \ac{sedt} solutions do not clearly state the key characteristics and advantages compared to existing security concepts.
Moreover, a consistent vocabulary and common view on the components cannot be established without a systematic classification of characteristics.
Overall, the current body of research gives rise to the following questions:
\begin{itemize}
    \item What are the promised advantages of \acp{sedt} for improving the security of \acp{cps}?
    \item How do \acp{sedt} differ from established security concepts and approaches?
    \item What are the characteristics of \acp{sedt} and how can they be evaluated?
    \item What are the current challenges and barriers faced by the \ac{cps} security community when developing \acp{sedt}?
\end{itemize}

The remainder of this article addresses each of the above questions to provide a source of reference for future research on \acp{sedt}.

\vspace{0.5em}
\begin{glossarybox}{Glossary of Terms \& Related Concepts}
\small
\textit{Cyber-physical system (CPS):} A CPS employs computing elements and interacts with the real world by means of sensors and actuators.\vspace{0.4em}\\
\textit{Cyber range:} A cyber range is a security testbed used for training and testing purposes.\vspace{0.4em}\\
\textit{Data-driven model:} A data-driven model is derived from previously collected data samples, for example, by using machine learning methods.\vspace{0.4em}\\
\textit{Deception technology:} Deception technologies, such as honeypots, are systems set up as decoys to detect and study cyberattacks.\vspace{0.4em}\\
\textit{Emulation:} An emulation mimics the inner workings of a system with the objective of substituting it in some analysis or test scenario.\vspace{0.4em}\\
\textit{Hardware-in-the-loop (HIL):} A HIL test setup consists of the real hardware of the embedded system under test and a simulated environment featuring the dynamics of the studied CPS.\vspace{0.4em}\\
\textit{Physical model:} A physical model (often also referred to as a plant model) is a mathematical representation of physical system behavior.\vspace{0.4em}\\
\textit{Simulation:} A simulation models the behavior of a system or phenomenon for analysis purposes.\vspace{0.4em}\\
\textit{Software-in-the-loop (SIL):} A SIL setup enables the testing of embedded software under simulated conditions as generated by the plant model.
\end{glossarybox}
\section{Promised Advantages of SEDTs}\label{sec:promises}

\begin{figure*}[ht!]
    \centering
    \begin{overpic}[scale=0.24]{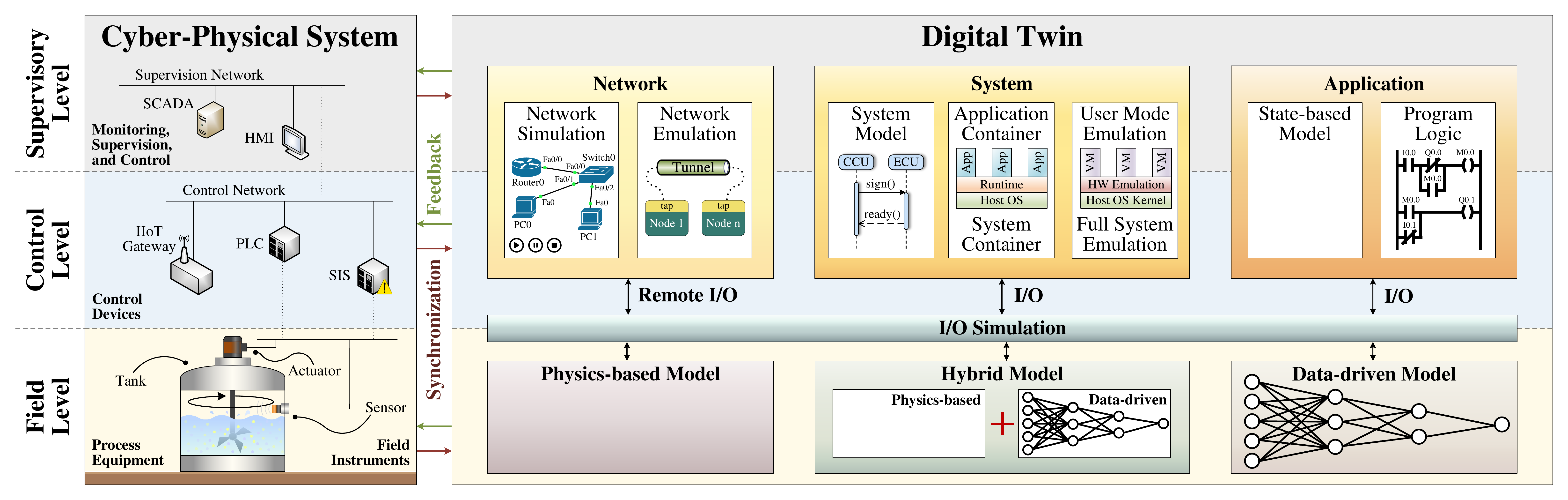}
        \put (34,5.85) {\scalebox{0.5}{$\rho\frac{\mathrm{D} \mathbf{u}}{\mathrm{D} t} = - \nabla p + \nabla \cdot \boldsymbol \tau + \rho\,\mathbf{g}$}}
        \put (35.2,4.2) {\scalebox{0.5}{$\boldsymbol \sigma = \lambda (\nabla\cdot\mathbf{u}) \mathbf I + 2 \mu \boldsymbol \varepsilon$}}
        \put (35.2,2.5) {\scalebox{0.5}{$\boldsymbol \tau = \mu \left(\nabla\mathbf{u} +  \nabla\mathbf{u} ^\mathrm{T}\right)$}}
        
        \put (54.5,5.2) {\scalebox{0.3}{$\rho\frac{\mathrm{D} \mathbf{u}}{\mathrm{D} t} = - \nabla p + \nabla \cdot \boldsymbol \tau + \rho\,\mathbf{g}$}}
        \put (55.2,4.3) {\scalebox{0.3}{$\boldsymbol \sigma = \lambda (\nabla\cdot\mathbf{u}) \mathbf I + 2 \mu \boldsymbol \varepsilon$}}
        \put (55.2,3.3) {\scalebox{0.3}{$\boldsymbol \tau = \mu \left(\nabla\mathbf{u} +  \nabla\mathbf{u} ^\mathrm{T}\right)$}}

        \put (80,15.8) {\scalebox{0.3}{
        \begin{tikzpicture}[shorten >=1pt,node distance=1.3cm,on grid,auto]
  \tikzstyle{every state}=[fill={rgb:black,1;white,10}]

    \node[state,initial]   (q_1)                    {$q_1$};
    \node[state,accepting] (q_2)  [below of=q_1]    {$q_2$};
    \node[state]           (q_3)  [below of=q_2]    {$q_3$};

    \path[->]
    (q_1) edge [loop right] node {0}    (   )
          edge [bend left]  node {1}    (q_2)
    (q_2) edge [bend left]  node {0}    (q_3)
          edge [loop right] node {1}    (   )
    (q_3) edge [bend left]  node {0,1}  (q_2);
\end{tikzpicture}
        }}
        
    \end{overpic}

    \vspace{-0.75em}
    \caption{Schematic structure of an automated industrial mixing process as an example of a \ac{cps} and its corresponding \ac{dt}.
    Note: A concrete \ac{dt} implementation may comprise a mix of the elements presented (depending on the security use case of interest) and the labels used in the illustration only indicate the general scope (e.g., control logic, SCADA software, and HMI software are all subsumed under ``program logic'').
    Furthermore, real systems and peripheral devices may also be connected to the \ac{dt} to address certain constraints such as emulator limitations.
    }
    \label{fig:use_cases}
\end{figure*}
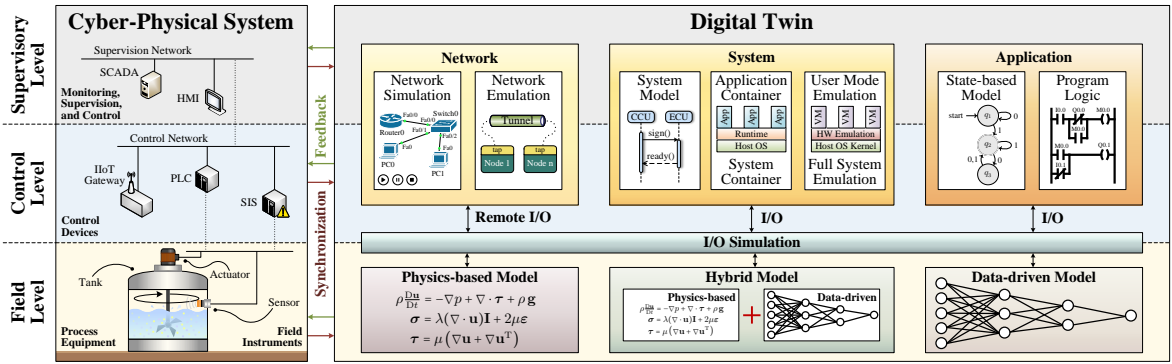

Fig.~\ref{fig:use_cases} illustrates potential elements of a \ac{dt} and its physical counterpart using the example of an industrial mixing system, which is part of an \ac{ics} employed at a chemical production site.
As shown, a \ac{dt} can be implemented by means of simulations, emulations, and data-driven models or a combination thereof.
Since the concrete implementation of the \ac{dt} depends on the use case of interest, Fig.~\ref{fig:use_cases} displays merely a collection of components that may appear in some form or another.

\acp{sedt} that are specifically designed to improve \ac{cps} security have various applications that boil down to one overarching benefit:
they allow users to gain a deeper understanding of the past, present, or future system behavior without the risk of causing operational disruption or physical damage.

\subsection{Security Use Cases of SEDTs}

In the following, we briefly describe those cybersecurity application areas where we consider that the usage of an \ac{sedt} yields the greatest benefit in the context of \ac{cps} security.

\noindent\textbf{Security Analysis and Risk Assessment.}
\acp{sedt} can benefit security analyses, as attack scenarios (including potential cascading and mutually amplifying effects) can be explored without affecting the normal operation of the real system.
As such, they enable users to explore hypothetical scenarios featuring threats and countermeasures on the virtual network, system, and application layers, while the resulting negative and positive effects can be observed on the field level via the integrated physics-based or data-driven model(s).
Moreover, an \ac{sedt} can assist in estimating the loss probability by determining the attacker's success paths.
It can also be used to expose experts and decision makers to worst-case scenarios when forming their judgments on loss severity.

\noindent\textbf{Security Testing and Certification.}
An \ac{sedt}, which includes relevant security mechanisms of a real system and can emulate behavior under real attacks, allows security analysts to routinely perform penetration testing and other forms of security testing without affecting the real systems.
Furthermore, such \acp{sedt} might be used to support security-~and safety-related certification activities by acting as a surrogate to demonstrate the \ac{cps}'s robustness against adverse events.

\noindent\textbf{Training.}
Using \acp{sedt} for training might offer flexibility and a sufficient level of realism, possibly making security exercises more effective and rewarding for the trainee.
\acp{sedt} have the added benefit of showing trainees the direct effects of security controls on a virtual representation of the system under consideration (e.g., via full system and network emulation).

\noindent\textbf{Forensics.}
\acp{sedt} offer the ability to analyze events without the risk of tampering with the real systems and available evidence~\cite{Dietz2020a}.
Furthermore, they provide a wealth of information (e.g., trace data, execution history, states of the physical system) to support forensic analysis and allow for preserving systems beyond their lifetime.
As a result, an incident could be discovered and investigated (with limitations) after the system has been modified (or even decommissioned).

\noindent\textbf{Intrusion Detection.}
Comparing the behavior of an \ac{sedt} with that of its physical counterpart further opens up the possibility of detecting intrusions.
Behavior-specification-based \acp{ids} rest upon a formal specification of legitimate behavior, which typically requires significant effort to create~\cite{Mitchell2014}.
An \ac{sedt}, which is an instantiation of a specification, does not ease this burden, especially since its scope is extended to cover the network, system, logic, and physics layers.
However, the multi-layered feature of an \ac{sedt} yields more audit material than existing \acp{ids} of this class, which typically collect either host-based or network-based data.
In the context of behavior-based intrusion detection, \acp{sedt} can also be used for data generation purposes as an alternative to gathering data from the real \ac{cps}.
This will not only provide greater flexibility in data acquisition but also ensure that potential intrusions are not already present in the training data (in case supervised learning is applied).
\acp{sedt} can also illustrate anomalous behavior by reacting to simulated cyberattacks and common system faults to complement data reflecting normal behavior.

\noindent\textbf{Response.}
Once an intrusion has been detected, \acp{sedt} can help to perform root-cause analysis and support the planning of a reactive response in order to minimize the attack impact and recover from the effects of a compromise.
This can be achieved by simulating similar threats against \acp{sedt} and testing possible countermeasures to assess their effectiveness as well as their effects on the physical process.

\noindent\textbf{Deception.}
The components developed in the course of \ac{sedt} creation might be reused to run a honeypot alongside other systems as part of the real \ac{cps}.
Since an \ac{sedt} is normally designed to closely replicate the behavior of its physical counterpart, its application as a honeypot could yield a high level of interaction and realism.

\noindent\textbf{Patch Management.}
\acp{sedt} can support patch management by providing the means to test patches on virtual replicas without disrupting or endangering real systems~\cite{Holmes2021}.

\subsection{Key Differences from Existing Security Approaches}
The previously described security applications of \acp{sedt} are inspired by existing concepts that are backed by an extensive body of research.
This naturally raises the question of what unique aspects are actually offered by \acp{sedt}.
We therefore highlight the anticipated capabilities of \acp{sedt} in comparison to established security approaches.

\noindent \textbf{Security Testbeds and Cyber Ranges.} \acp{sedt}, as purely software-based solutions, provide a cost-efficient alternative to cyber ranges that integrate physical elements, provided that proper tools to generate and operate them are available.
Specifically, the procurement and provisioning for hardware-based environments typically entail considerable costs.
Furthermore, even after ramp-up, (semi-)physical testbeds demand manual effort to manage devices (e.g., modify setups, tear down after test execution).
Uncontrolled failures may also lead to damaged equipment, incurring additional expenses.

\noindent \textbf{Software-in-the-loop (SIL).}
We view the \ac{sedt} as a more capable solution that goes beyond functional testing of a \ac{cps} as it provides the means to study the behavior of the system under unintended adverse conditions while considering feedback from the real-world environment.

\noindent \textbf{Security Modeling Tools.}
\acp{sedt} may transform the fragmented security solution landscape, which is characterized by information silos that emerged with the proliferation of individual, custom-built (sub-)models for security analysis.
The assessment scope of security analysis tools is typically limited to specific parts of the \ac{cps}, and the employed information models are often stored in proprietary formats, leading to isolated consideration of security aspects.
The \ac{sedt}, as a unified virtual representation of the \ac{cps}, connects multiple models to support security analyses from different perspectives.
This advantageous trait also enables ``security by design'' for \acp{cps}:
an \ac{sedt} can exist already in the early phases of the \ac{cps} lifecycle, even before the real system is built, and thereby inform engineers about security-relevant issues.
As the \ac{cps} evolves throughout its lifecycle, the \ac{sedt} functions as a digital companion to support continuous security upgrades.

\noindent \textbf{Data-Driven Models.}
It is worth noting that the \ac{dt}, in general, is often understood as a data-driven model constructed with data collected from real-world objects~\cite{Negri2017, Jones2020}.
In contrast to this position, we adopt the ``classical'' perspective of the \ac{dt} concept, which does not restrict \acp{dt} to pure data-driven models but rather embraces an integrated approach that combines them with physics-based models and system models (as originally envisaged by NASA~\cite{Shafto2012}).
While \acp{sedt} certainly build upon historical and real-time data coming from the \ac{cps}, their method of construction is typically not limited to machine learning algorithms and may also include system emulation, network simulation/emulation, models of control logic, and physics-based models (cf.~Fig.~\ref{fig:use_cases}).

\noindent \textbf{Deception Technology.}
When \acp{sedt} are specifically designed to function as a form of deception, there will be a high degree of technological overlap with high-interactive honeypot systems, as the \acp{sedt} should be barely distinguishable from real systems in order to deceive attackers.
One advantage of \acp{sedt} is that they may facilitate the combination of deception and \ac{mtd} approaches by supporting dynamic changes to the virtual environment, aiming to interfere with the attacker's efforts to identify decoys.
However, an \ac{sedt} used as a decoy should not have a bidirectional connection (or unidirectional connection DT $\rightarrow$ CPS) to avoid any negative ``spillover'' from an attack.
Moreover, precaution must be taken when using \acp{sedt} as decoys since they may disclose valuable information about the actual systems.

\section{Characteristics of SEDTs}\label{sec:characteristics}

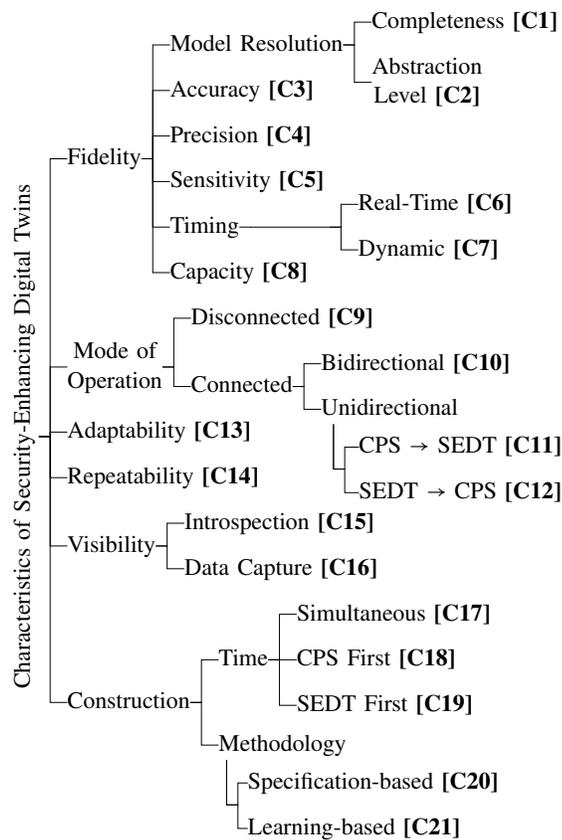
\begin{figure} 
    \centering  
     \scalebox{0.85}{
\begin{forest}
    for tree={
      grow=east,
      parent anchor=east,
      child anchor=west,
      align=center,
      l=1em, l sep+=0em,anchor=base west,inner sep=0.1pt,outer sep=0pt,
       edge path={
                \noexpand\path [draw, \forestoption{edge}] (!u.parent anchor) -- +(5pt,0) |- (.child anchor)\forestoption{edge label};
            },
      for root={
        parent anchor=east,
      },
    }
    [Characteristics of Security-Enhancing Digital Twins,rotate=90,child anchor=north, parent anchor=south, anchor=center
      [Construction,s sep=-2mm,
        [Methodology \phantom{XZY}, parent anchor=-170,
            [Learning-based {\textbf{[C21]}},edge path={
                \noexpand\path [draw, \forestoption{edge}] (!u.parent anchor) |- +(5pt,-20pt) |- (.child anchor)\forestoption{edge label};
            },before drawing tree={x-=82,y-=27pt}]
            [Specification-based {\textbf{[C20]}},edge path={
                \noexpand\path [draw, \forestoption{edge}] (!u.parent anchor) |- +(5pt,-20pt) |- (.child anchor)\forestoption{edge label};
            },before drawing tree={x-=82,y-=27pt}]
        ]
        [Time
            [SEDT First {\textbf{[C19]}}]
            [CPS First {\textbf{[C18]}}]
            [Simultaneous {\textbf{[C17]}}]
        ]
      ]
      [Visibility
        [Data Capture {\textbf{[C16]}}]
        [Introspection {\textbf{[C15]}}]
      ]
      [Repeatability {\textbf{[C14]}}]
      [Adaptability {\textbf{[C13]}}]
      [Mode of\\Operation,
        [Connected
            [Unidirectional \phantom{Flow}, parent anchor=-170
                [SEDT $\rightarrow$ CPS {\textbf{[C12]}},edge path={
                \noexpand\path [draw, \forestoption{edge}] (!u.parent anchor) |- +(5pt,-20pt) |- (.child anchor)\forestoption{edge label};
            },before drawing tree={x-=82,y-=27pt}]
            [CPS $\rightarrow$ SEDT {\textbf{[C11]}},edge path={
                \noexpand\path [draw, \forestoption{edge}] (!u.parent anchor) |- +(5pt,-20pt) |- (.child anchor)\forestoption{edge label};
            },before drawing tree={x-=82,y-=27pt}]
            ]
            [Bidirectional {\textbf{[C10]}}]
        ]
        [Disconnected {\textbf{[C9]}}]
      ]
      [Fidelity
          [Capacity {\textbf{[C8]}}]
        [Timing,
            [,coordinate,l=2.5cm,yshift=0.91mm
                        [Dynamic {\textbf{[C7]}}]
                        [Real-Time {\textbf{[C6]}}]
            ]
          ]
          [Sensitivity {\textbf{[C5]}}]
          [Precision {\textbf{[C4]}}]
          [Accuracy {\textbf{[C3]}}]
        [Model Resolution,
                [Abstraction\\Level {\textbf{[C2]}}]
                [Completeness {\textbf{[C1]}}]
          ]
     ]
    ]
  \end{forest}

}
     \vspace{-1.6em} 
    \caption{Characteristics of security-enhancing \aclp{dt}.}
    \label{fig:characteristics}
\end{figure}

Existing scholarly and professional publications on the characteristics of \acp{dt} lack a thorough consideration of the requirements needed to implement security-enhancing use cases.
To address this gap, we have co-organized a Dagstuhl seminar~\cite{Mora2022} on \acp{dt} for \ac{cps} Security.
The taxonomy presented in Figure~\ref{fig:characteristics} was derived through workshop-style discussions on the requirements of the described security-related purposes in addition to a systematic analysis of the general characteristics of \acp{dt} (e.g.,~\cite{Jones2020, Valk2020, Budiardjo2021}).
The introduced taxonomy describes the characteristics of \acp{sedt} in a structured manner and enables the classification of future \ac{dt}-based security solution proposals.
Based on this, we offer suggestions for qualitative and quantitative indicators to evaluate the proposed characteristics, aiming to provide a starting point for a fair comparison among \ac{sedt} solutions.

\subsection*{Fidelity [C1]--[C8]}
The term \emph{fidelity} is loosely used in the modeling and simulation community to refer to the ``level of detail'' of a simulation, yet a widely agreed and clear definition is missing~\cite{Liu2009a}.
In the context of \acp{sedt}, the understanding of this term is correspondingly vague:
the fidelity of an \ac{sedt} refers to how closely it resembles its physical counterpart.
We argue that the lack of clarity about fidelity hampers the adoption of the \ac{dt} concept, as misunderstandings may already arise during requirements engineering and propagate to subsequent phases.
Furthermore, this obscure definition inhibits the measurement of fidelity, which is necessary to assess the suitability for purpose~\cite{Liu2009a}.
To narrow the room for interpretation, Roza et al.~\cite{Roza2000} proposed fidelity concepts that decompose the notion of fidelity into more concrete elements, and thereby provide a clearer sense of this term.

Before we introduce our notion of \ac{dt} fidelity, we want to direct the reader's attention to the following important issues raised by Roza et al.~\cite{Roza2000}:
First, fidelity is measured by comparing the virtual replica to the perceived real-world counterpart. 
Naturally, the perception of reality is incomplete at best and flawed at worst.
Second, comparing observable properties of the real system can only be indicative of how well the data points are replicated (by the \ac{dt}), not how well the actual behavior is reflected.
Third, such virtual replicas may be employed at a stage where the real system is still in development or completely non-existent, making data collection for fidelity measurement infeasible. 
Roza et al.~\cite{Roza2000} addressed these issues by introducing the concept of a \emph{fidelity referent}, which can be understood as a description of knowledge that approximates reality.
Such specifications offer a pragmatic basis for fidelity measurement.

In the following, we build upon the fidelity concepts of Roza et al.~\cite{Roza2000} and adapt them to the context of \acp{dt} (and, more specifically, \acp{sedt}):

    \noindent \textbf{Model Resolution [C1--C2]} comprises the completeness and abstraction level of the \ac{sedt}.
    
    Completeness~[C1] is a measure of how exhaustive in breadth the \ac{cps} is virtually replicated.
    In the system-of-systems context, this characteristic indicates the extent of the \ac{cps} for which \acp{sedt} exist.
    For instance, in an isolated view that is strictly limited to computer systems, a complete \ac{sedt} of an \ac{ics} would be composed of a set of \acp{sedt}, where each member corresponds to one system employed as part of the real \ac{ics}.
    Due to the physical dimension, defining full completeness, let alone achieving it, is infeasible.
    Nevertheless, the completeness of an \ac{sedt} (also taking the physical system into account) can be evaluated with respect to a referent that specifies the physical properties that would have to be modeled for achieving the desired security purpose.
    
    The degree of abstraction of an \ac{sedt}~[C2] is influenced by its ``depth'' (i.e., detail).
    The spectrum of abstraction relates to several levels in which \acp{cps} function. 
    On the computer system level, this may take the form of a simulated model of computation (e.g., Petri net), a simulated system model (e.g., based on a SysML model), or full system emulation (e.g., via QEMU).
    Similarly, on the network level, simulating communication networks will yield a higher degree of abstraction than fully emulating the network stack.
    Opting for reduced computational complexity in lieu of lower abstraction by employing a reduced-order model as an approximation to the full-order model is an example for the physical level.
    
    \noindent \textit{Relevance to Security Use Cases.}
    The required completeness~[C1] of the \ac{sedt} primarily depends upon the purpose and scope of the security use case(s), which determine the computer systems, physical processes, network infrastructure, etc.~that need to be covered.
    
    Similarly, the requirements relating to the abstraction level~[C2] strongly depend on the chosen security use case, as well as the scope and purpose of evaluation.
    For instance, determining attack scenarios and assessing their effects may not require a low level of abstraction since attack simulations on system models are presumably sufficient to understand potential attack paths and consequences.
    On the other hand, conducting threat hunting or investigating the attack paths in play during an incident in order to find the next pivot action that is likely to be taken by adversaries will necessitate a more detailed emulation of systems. 
    In addition, \acp{sedt} deployed as decoys must provide a high level of interaction, which is governed by the degree of abstraction, in order to avoid identification as honeypots.
    However, low abstraction may not necessarily imply equivalent utility across all security use cases. 
    For example, an \ac{sedt} with low abstraction used as a basis for an \ac{ids} may not directly lead to increased detection performance.
    The reason for this is that less detail (i.e., higher abstraction level) may result in greater robustness in terms of sensitivity and fewer synchronization errors.
    
    \noindent \textit{Indicator Considerations.}
    Completeness~[C1] can be measured as the proportion of the covered \ac{cps} components and physical properties of the real systems, which are required for the intended security purpose.
    On the other hand, the abstraction level~[C2] can be measured by gauging which details have been left out (or generalized) with respect to a referent.

    \noindent \textit{Exemplary Sources.}
    Suggested sources for creating a referent are engineering artifacts detailing the plant topology (i.e., structure of system resources and communication networks) to cover the digital elements of the \ac{cps}, whereas information about the physical part can be obtained from the specification of theoretical knowledge about the physical phenomenon and empirical data collected from the physical system.

    \noindent \textbf{Accuracy [C3]} describes how close an observed property of an \ac{sedt} is to the observed property of the real system (or the \emph{true} value, if it is known).
    An \ac{sedt} with high accuracy would produce the same outputs as the \ac{cps} when presented with the same inputs and environmental conditions.
    If accuracy decreases below a certain threshold, the \ac{sedt} may not be considered a twin anymore.
    
    \noindent \textit{Relevance to Security Use Cases.}
    Using an \ac{sedt} as a form of deception is the only security purpose where accuracy can be sacrificed for practicality.
    High levels of accuracy are a must-have requirement for the other security application areas, where defenders consuming the outputs produced by \ac{sedt} heavily rely on the accuracy of results to make well-informed decisions.
    For example, the \ac{ids} performance (i.e., measured in terms of false positive/negative rates) is heavily influenced by the accuracy of the underlying \ac{sedt}.
    
    \noindent \textit{Indicator Considerations.}
    Accuracy can be expressed in quantitative terms using classical error measures, such as absolute error and relative error.
    
    \noindent \textit{Exemplary Sources.} 
    A referent may be sourced from empirical knowledge or from subject matter experts who specify referent data points.
    
    \noindent \textbf{Precision [C4]} refers to the degree of exactness, in terms of the resolution or granularity of representation, of the outputs or results produced by an \ac{sedt}.
    The concrete manifestation of precision depends on the type and format of the considered output values.
    In the most basic case, this can be the arithmetic precision of a numeric value.
    Limited precision may be caused by the \ac{sedt}'s level of abstraction, inherent shortcomings of its implementation, or even deliberately accepted with the intention of increasing efficiency.
    For instance, round-off errors are a natural consequence of finite arithmetic applied in numerical computation, whereas simplifications in mathematical calculations are made to improve the performance, possibly at the cost of lower precision.
    
    \noindent \textit{Relevance to Security Use Cases.}
    Low-precision computation can have negative effects on the accuracy of the \ac{sedt}.
    In particular, less precise output values may introduce or amplify errors along the \ac{sedt} execution path, undermining the utility of analysis outcomes.
    However, for certain use cases (e.g., deception), the resulting accuracy drop may be acceptable.
    
    \noindent \textit{Indicator Considerations.}
    In numerical analysis, a quantitative indicator of precision is given by the total number of significant digits.
    For non-numeric outputs, qualitative indicators can be established and checked against referent knowledge.
    For example, consider precision as it pertains to network communication:
    simulating every individual network packet would yield higher granularity than flow-level simulation---yet at the cost of greater computational effort.
    
    \noindent \textit{Exemplary Sources.}
    The precision of the \ac{cps}'s computer systems can serve as a baseline for the corresponding models used as part of the \acp{sedt}.
    Approximating physics necessitates careful consideration of various parameters, such as modeling approach, phenomena to be studied, and computing environment.
    
    \noindent \textbf{Sensitivity [C5]} indicates how an \ac{sedt}'s behavior is affected by internal or external input inaccuracies.
    High sensitivity can negatively affect the execution of \acp{sedt} in all modes of operation.
    Normally, the \ac{sedt} setup consists of multiple instances that interact with each other; hence, output errors may propagate in an uncontrolled way and accumulate throughout the execution process.
    Furthermore, additional sources of error can emerge when \acp{sedt} are synchronized with their physical counterparts, as the real-world data is often noisy and may be incomplete.
    
    \noindent \textit{Relevance to Security Use Cases.}
    Sensitivity has a direct influence on those security use cases where interaction among \acp{sedt} exists (e.g., via I/Os) or \acp{sedt} are synchronized with their corresponding real systems.
    
    \noindent \textit{Indicator Considerations.}
    Sensitivity can be studied in a standalone setting or with respect to the \ac{cps} (or referent).
    In the former case, general practices can be borrowed from the field of sensitivity analysis, whereas indicators of the latter category are obtained by measuring the error in output values in terms of the deviation from the benign, real system in different modes of operation (e.g., accumulated error through the execution of one or multiple \acp{sedt}).
    
    \noindent \textit{Exemplary Sources.}
    The input/output behavior of the \ac{cps}, provided that it exists and proper test conditions can be established, may serve as a primary source for the specification of relevant referent knowledge.
    
    \noindent \textbf{Timing [C6]--[C7]} expresses how the state of an \ac{sedt} advances in relation to its physical counterpart.
    This characteristic can be further subdivided into timing configurations of \acp{sedt}, viz., real-time and dynamic.
    To run an \ac{sedt} synchronous to the corresponding real system, real-time support~[C6] is required, meaning that it is executed in discrete time with a constant step size sufficiently approximating the continuous behavior of the physical counterpart.
    In order to achieve this, the \ac{sedt} should advance at least at the same rate as the computationally-enabled components of the \ac{cps}.
    The possibility to (dynamically) accelerate or decelerate the execution of the \ac{sedt}~[C7] can also be important for certain use cases, such as predictive analysis.
    
    \noindent \textit{Relevance to Security Use Cases.}
    As indicated above, real-time support is mandatory for those security use cases that require the \ac{sedt} to be synchronized with its counterpart (e.g., intrusion detection, forensics) or if it must perform at the same rate as the actual system (e.g., security testing, training, deception, patch management).
    The support of dynamic temporal resolution, on the other hand, may enable a forward-looking perspective on how the state of the corresponding system might evolve over time after a certain activity has been performed (e.g., implementation of countermeasures to respond to a detected intrusion).
    
    \noindent \textit{Indicator Considerations.}
    The real-time characteristic~[C6] of an \ac{sedt} can be classified and measured in the same way as real-time systems. 
    Timing analysis may be conducted to assess, inter alia, latency and jitter, aiming to determine if it can be guaranteed that deadlines are met.
    Indicators for dynamic timing~[C7] can be both qualitative and quantitative:
    \begin{itemize}[leftmargin=*]
        \item \underline{Feature-wise}, to describe key qualities of the functionality provided (e.g., time resolution and speed adjustments, single stepping, conditional breakpoints).
        \item \underline{Accuracy-wise}, as a proxy measure, to determine how accurate the results of the execution with variable control over steps are.
    \end{itemize}
    
    \noindent \textit{Exemplary Sources.}
    Information for the referent may come from specified time requirements with respect to the task execution of control devices being used in the \ac{cps} (e.g., cycle times, jitter tolerance).
    
    \noindent \textbf{Capacity [C8]} relates to the performance of an \ac{sedt} implementation.
    This characteristic is influenced by the hardware and software used to run \acp{sedt} and dictates, for example, the number of instances that can be executed on a given node at the same time.
    
    \noindent \textit{Relevance to Security Use Cases.}
    As is the case with completeness~[C1], capacity requirements are mainly driven by the security use case of interest and the \ac{cps} at hand (e.g., runtime overhead caused by virtually replicating the control level and field level).
    
    \noindent \textit{Indicator Considerations.}
    Capacity can be determined by measuring the performance of the (hard-~and) software components of the \ac{sedt} implementation.
    Analyzing algorithmic efficiency and virtualization overhead are two examples of how this characteristic can be assessed.
    
\subsection*{Mode of Operation [C9]--[C12]}
We distinguish between different modes of operation based on the communication between the \ac{cps} and \acp{sedt}.

    \noindent \textbf{Disconnected~[C9]:}
    As the name implies, no data flow exists between the physical (i.e., real-world) environment and the virtual environment.
    This mode essentially resembles the conventional runtime options of simulations (batch or interactive), where \acp{sedt} are executed with a set of initial parameter settings and run independently of their physical counterparts.
    Thus, state changes of the real system are not reflected in the corresponding \ac{sedt}.
    
    \noindent \textit{Relevance to Security Use Cases.}
    For certain use cases, such as security testing or training, it may be mandatory that the \ac{sedt} is disconnected from the real system to avoid any accidental disturbance; yet, it may still be beneficial to initialize the \ac{sedt} with a state previously observed in the real world to create a suitable test or training environment.
    
    \noindent \textbf{Connected~[C10]--[C12]:}
    A physical entity can be linked to its digital twin via a unidirectional~[C11]--[C12] or even bidirectional~[C10] connection.
    The unidirectional category can be further subdivided into two groups on the basis of where data flows originate.
    
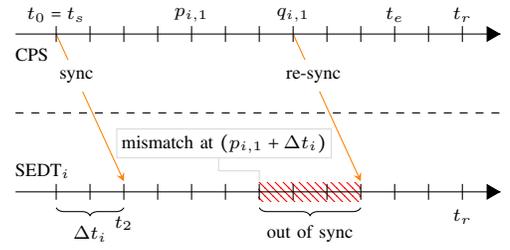
\begin{figure} 
    \centering  
\begin{tikzpicture}[y=.14cm, x=.36cm]
    	\draw (1,0) -- coordinate (x axis dt) (18.9,0);
    	\draw (1,15) -- coordinate (x axis cps) (18.9,15);

        \node at (1.6,13) {\scriptsize CPS};
        \node at (2,2) {\scriptsize $\text{SEDT}_i$};

        \begin{pgfonlayer}{bg}
            \draw[dashed] (1,7.5) -- coordinate (x axis boundary) (18.9,7.5);

            \draw[decorate,decoration={brace,amplitude=3pt,mirror}] 
            (2.5,-1.5) -- (5,-1.5); 
            
             \node [rectangle,fill,minimum width=1.35cm,minimum height=0.0cm,pattern=north west lines,pattern color=red] at (11.86, 0) {};
             
        \draw[decorate,decoration={brace,amplitude=3pt,mirror}] 
            (10,-1.5) -- (13.75,-1.5); 
            
        \draw [-stealth,color=orange](2.5,15) -- (5,1);
        
        \draw [-stealth,color=orange](11.25,15) -- (13.75,1);
            
        \node at (3.75,-4){\scriptsize$\Delta t_i$};
        
        \node at (11.875,-4){\scriptsize out of sync};

        \end{pgfonlayer}
        
        \node[fill=white] at (12,11){\scriptsize re-sync};
        
        \node[fill=white] at (3.25,11){\scriptsize sync};

        \node[draw, inner sep=0.065cm, color=gray!30, text=black] at (8.7,4.6) {\scriptsize$\text{mismatch at } (p_{i,1} + \Delta t_i)$};
        
        \draw[color=gray!30] (10,1) -- (10,2) -- (8.5, 2) -- (8.5, 3.2);


        \draw ([yshift=3pt]2.5,0) -- ([yshift=-3pt]2.5,0);
    	\draw ([yshift=3pt]2.5,15) -- ([yshift=-3pt]2.5,15)  node[anchor=south,yshift=4pt] {\scriptsize$t_0 = t_s$};
    	
    	\draw ([yshift=3pt]3.75,0) -- ([yshift=-3pt]3.75,0);
    	\draw ([yshift=3pt]3.75,15) -- ([yshift=-3pt]3.75,15);
    	
    	\draw ([yshift=3pt]5,0) -- ([yshift=-3pt]5,0) node[yshift=-2.5pt,anchor=north] {\scriptsize$t_2$};
    	\draw ([yshift=3pt]5,15) -- ([yshift=-3pt]5,15);
    	
    	\draw ([yshift=3pt]6.25,0) -- ([yshift=-3pt]6.25,0);
    	\draw ([yshift=3pt]6.25,15) -- ([yshift=-3pt]6.25,15);
    	
    	\draw ([yshift=3pt]7.5,0) -- ([yshift=-3pt]7.5,0);
    	\draw ([yshift=3pt]7.5,15) -- ([yshift=-3pt]7.5,15) node[anchor=south,yshift=4pt] {\scriptsize$p_{i,1}$};
    	
    	\draw ([yshift=3pt]8.75,0) -- ([yshift=-3pt]8.75,0);
    	\draw ([yshift=3pt]8.75,15) -- ([yshift=-3pt]8.75,15);
    		
    	\draw ([yshift=3pt]10,0) -- ([yshift=-3pt]10,0);
    	\draw ([yshift=3pt]10,15) -- ([yshift=-3pt]10,15);

        \draw ([yshift=3pt]11.25,0) -- ([yshift=-3pt]11.25,0);
        \draw ([yshift=3pt]11.25,15) -- ([yshift=-3pt]11.25,15) node[anchor=south,yshift=4pt] {\scriptsize$q_{i,1}$};
    	
    	\draw ([yshift=3pt]12.5,0) -- ([yshift=-3pt]12.5,0);
    	\draw ([yshift=3pt]12.5,15) -- ([yshift=-3pt]12.5,15);
    	
    	\draw ([yshift=3pt]13.75,0) -- ([yshift=-3pt]13.75,0);
    	\draw ([yshift=3pt]13.75,15) -- ([yshift=-3pt]13.75,15);
    	
    	\draw ([yshift=3pt]15,0) -- ([yshift=-3pt]15,0);
    	\draw ([yshift=3pt]15,15) -- ([yshift=-3pt]15,15) node[anchor=south,yshift=4pt] {\scriptsize$t_e$};
    	
    	\draw ([yshift=3pt]16.25,0) -- ([yshift=-3pt]16.25,0);
    	\draw ([yshift=3pt]16.25,15) -- ([yshift=-3pt]16.25,15);
    	
    	\draw ([yshift=3pt]17.5,0) -- ([yshift=-3pt]17.5,0) node[anchor=north] {\scriptsize$t_r$};
    	\draw ([yshift=3pt]17.5,15) -- ([yshift=-3pt]17.5,15) node[anchor=south,yshift=4pt] {\scriptsize$t_r$};
    	
        \draw [-triangle 60] (18.9,0) -- (18.91,0);
        \draw [-triangle 60] (18.9,15) -- (18.91,15);

    \end{tikzpicture}
\vspace{-0.75em}
\caption{
Example of a synchronization session (assuming a discrete approximation of time, where $T = \{0,\dots,r\}$ and $t_k \in T$).
$\text{SEDT}_i$ is synchronized over the time span $t_s$ to $t_e$ and follows the states of the respective real system, which is part of a \ac{cps}, with delay $\Delta t_i = 2 \text{ time intervals}$.
The synchronization session starts at $t_0$ and ends at $t_{10}$, with one out-of-sync region in between (3 time intervals).
$p_{ij}$ and $q_{ij}$ are the start and end of $\text{SEDT}_i$ being out of sync (relative to the time scale of the \ac{cps}).}
    \label{fig:synchronization}
\end{figure}

    Data flows originating from the real-world environment~[C11] (or~[C10]) are used to synchronize the \ac{sedt} with its physical counterpart, ensuring that it replicates the current state with a certain delay.
    Such synchronization is implemented by means of a state replication mechanism that collects data from the \ac{cps} either in a passive~\cite{Eckhart2018a} (e.g., network traces, system logs, measurements of the plant's output) or active manner~\cite{Gehrmann2020} (e.g., polling).
    Note that running \acp{sedt} in this mode does not necessarily mean that the state representation within \acp{sedt} must \emph{exactly} match the one of the real systems since they might combine several states into one abstract state.
    Furthermore, state replication mechanisms can be categorized according to the scope of synchronization, viz., partial sync and full sync (i.e., selective synchronization of some \acp{sedt} or all of them).
    Fig.~\ref{fig:synchronization} visualizes the core ideas of \ac{sedt} synchronization.

    An \ac{sedt} may also have a direct data link to its physical counterpart~[C12] (or~[C10]) to implement a feedback mechanism.
    For instance, if the evaluated (reactive) strategy to counter current threats has the desired effect on the \ac{sedt}, the re-configurations and countermeasures applied in the virtual environment can be carried over to the \ac{cps}.
    In a similar vein, proactive responses can be initiated when new security weaknesses have been revealed, even if the \ac{cps} is already in operation.
    We recognize that different tiers of backflow to the \ac{cps} exist, which indicate the achieved level of automation in terms of how the \ac{cps} uses this feedback.
    
    \noindent \textit{Relevance to Security Use Cases.}
    A connection between the \ac{cps} and the \ac{sedt} is required in one of the following cases:
    \begin{enumerate*}[label=(\roman*)]
        \item to base the intended security measure implemented by means of an \ac{sedt} on a current state of the \ac{cps} (e.g., security analysis, forensics, intrusion detection);
        \item to directly transfer the changes previously tested on an \ac{sedt} to the respective real system (e.g., rolling out recovery procedures).
    \end{enumerate*}

\noindent \textit{Indicator Considerations.}
The performance of state replication methods can be quantitatively assessed in various ways:
For instance, by calculating the mean time between state mismatches (higher is better) and mean state replication delay across all synchronized \acp{sedt} (lower is better).
In addition to these quantitative indicators, we suggest qualifiers to describe the degree of autonomy achieved when an \ac{sedt} provides input to the \ac{cps}:
\begin{itemize}[leftmargin=*]
    \item \underline{Manual.} The results obtained from the \ac{sedt} are only for human consumption and require a manual process to make use of them (e.g., \ac{ids} alerts users upon detection of malicious activity).
    \item \underline{Cooperative.} The \ac{sedt} automatically suggests (tested) changes but depends on human interaction (e.g., manual approval) to deploy them to the production environment.
    \item \underline{Autonomous.} No human intervention is required at this level, as the \ac{sedt} identifies, tests, and fixes issues autonomously such that the \ac{cps} can fully adapt to the dynamic environment.
\end{itemize}

\subsection*{Adaptability [C13]}
This characteristic refers to the possibility of changing components and configurations of the \ac{sedt}.
For instance, potential changes to the \ac{sedt} may relate to the peripherals of the emulated system, network configuration, and simulation parameters.

\noindent \textit{Relevance to Security Use Cases.}
An \ac{sedt} with adaptability capabilities could facilitate security analysis, security testing, and training by enabling quick iterations with continuous configuration changes, thereby fostering frequent feedback.

\noindent \textit{Indicator Considerations.}
Qualitative indicators can be established on different levels:
\begin{enumerate*}[label=(\roman*)]
\item industrial domain level (e.g., an \ac{sedt} originally employed for a wastewater treatment system is reconfigurable to also support natural gas power plants),
\item \ac{cps} architecture level (e.g., adaptations of \acp{sedt} in a system-of-systems context),
\item component level (e.g., parameters and structure of the models employed in an \ac{sedt}),
\item runtime level (e.g., modes of operation),
\item construction level (how the \ac{sedt} evolves with respect to the lifecycle of the \ac{cps}), and
\item feedback level (e.g., static or dynamic reconfiguration of \acp{sedt} after test cases failed).
\end{enumerate*}

\subsection*{Repeatability [C14]}
Two identical \acp{sedt} that receive equivalent external inputs should produce equal states and outputs, given the same initial conditions.
This key property ensures that resetting an \ac{sedt} to a previous (known) state is reasonable.
For instance, users might want to save and return to a specific point in time in order to repeat a scenario or to try a different investigation procedure from the same starting point.
To do this, the ability to create snapshots (i.e., capture the history of states of a system at a point in time) is critical.
It follows that non-deterministic behavior in simulations integrated into an \ac{sedt} inhibits repeatability.

\noindent \textit{Relevance to Security Use Cases.}
Repeatability is essential in cases where the results of multiple \ac{sedt} runs must be comparable.
Exploring alternative scenarios with the same \ac{sedt}, restoring past states to repeat tests, and providing a controlled setup to compare the consequences of deployed patches are a few examples of security purposes where repeatability is strictly required.

\noindent \textit{Indicator Considerations.}
An indicator for repeatability is obtained by measuring how close a result gained through an \ac{sedt} is compared to another result produced by the same \ac{sedt} in a repeated experiment under unchanged conditions.

\subsection*{Visibility [C15]--[C16]}

\acp{sedt} are of value to operators of \acp{cps} only if the encapsulated states and produced results are readily accessible. 
The following two characteristics dominate the visibility of \acp{sedt}.

    \noindent \textbf{Introspection [C15]:}
    This characteristic indicates the extent to which the state of an \ac{sedt} can be analyzed and is, therefore, primarily driven by its fidelity.
    A higher level of introspection allows user interaction and provides access to intermediate results.
    
    \noindent \textit{Relevance to Security Use Cases.}
    \acp{sedt} can offer means to analyze the internal system state that is otherwise difficult to measure or simply not accessible from the real-world system.
    New knowledge of the system and its behavior can be extracted by exploring the \ac{sedt} without fear of real-world consequences, with greater convenience and expedience and fewer physical or logical obstacles. 
    
    \noindent \textit{Indicator Considerations.}
    The different manifestations of \acp{sedt} (cf.~Fig.~\ref{fig:use_cases}) span a broad spectrum of interaction levels that can be ranked using qualifiers.
    For example, at the upper end of the spectrum, users can examine the state of the virtual machine by debugging the guest operating system and inspecting the running processes of control applications.
    
    \noindent \textbf{Data Capture [C16]:}
    Data capture refers to the capability of the \ac{sedt} to monitor and record data during its execution.
    Depending on the use case, different types of data points can be of interest for each component of the \ac{sedt} (e.g., physical properties, application logs, network traces).
    We distinguish if the \ac{sedt} can capture the state of the entire system or only that of specific components.
    As with introspection, concerns may arise that relate to the fidelity and the resulting accuracy of the data produced.
    
    \noindent \textit{Relevance to Security Use Cases.}
    In general, the promise of the \ac{sedt} is to enable more comprehensive and easier data capture than what would be possible from the physical twin.
    For example, data capture is important to understand where manifestations of security risks occur (i.e., in which subsystem) and how they affect the physical process under control.
    
    \noindent \textit{Indicator Considerations.}
    Like introspection, the offered data capture features can be classified across all \ac{cps} layers in a qualitative manner.
    Examples include: Logging the state of virtual machines, collecting network traffic, acquiring debugging information of control applications, and recording simulation data.
    
\subsection*{Methodology and Time of Construction [C17]--[C21]}

The methodology of construction can describe how the \ac{sedt} and its parts are built. 
In certain cases, the \ac{sedt} is built before the physical counterpart~[C19]; in others, the \ac{sedt} is constructed after the physical counterpart (e.g., for already existing systems)~[C18], or both can even be developed simultaneously~[C17].
These characteristics are sometimes also grouped under the term ``time of creation''~\cite{Valk2020}.
In either of these scenarios, for a while only one instance could exist (the physical or the digital part), which could also affect other characteristics described in this section.

The chronology of the creation of the two parts also limits the methods and sources available to develop \acp{sedt}. 
They can be built from specifications~[C20], by learning~[C21] (observing the behavior of the \ac{cps}), or a combination thereof. 

\noindent \textit{Relevance to Security Use Cases.}
The methodology for building the \ac{sedt} can have ramifications when it is used for security. 
For example, building the \ac{sedt} entirely by learning from the \ac{cps} and then using it as a basis to generate data for creating a behavior-based \ac{ids} would lead to an unnecessary indirection.
In this case, it seems more fruitful (e.g., in terms of accuracy) to directly use the original data.
\section{Research and Engineering Challenges}\label{sec:challenges}

Despite the great potential of \acp{sedt} for security purposes, we are still in the infancy of this concept and face a series of notable challenges that need to be addressed to realize its benefits.

\noindent \textbf{\acs{cps} Emulation \& Virtualization Tooling.}
One severe limitation is the emulation of field and control devices, such as \acp{plc}, since their hardware and software components are often proprietary and closed-source.
However, the advent of \acp{plc} with embedded Linux and the growth of open-source initiatives could alleviate this situation.
Nevertheless, suitable system emulator targets will be required.

\noindent \textbf{Trade-Off Between Fidelity \& Cost.}
Another challenge is to balance the fidelity of an \ac{sedt} and the effort of implementation such that target use cases are sufficiently supported while costs and development time are kept under control.
Naturally, high-fidelity \acp{sedt} necessitate the integration of sophisticated physical models and a (near-)complete coverage of components via emulations, which requires considerable implementation effort or is simply infeasible. 
On the other hand, imperfections of models are inevitable; thus, care has to be taken that wrong assumptions or mismatches will not lead to a false sense of assurance and, subsequently, wrong decision-making.
One promising solution to keep the costs and time for approximating the physics manageable is to combine the conventional physics-based approach with data-driven modeling if approximating through learning is sufficient for the intended purpose.
It is also worth noting that there are additional factors, which may be interrelated with fidelity or cost, that need to be considered when implementing \acp{sedt}.
For instance, higher fidelity \acp{sedt} may generate significant network traffic, thereby potentially affecting network performance or even the \ac{cps} itself (e.g., in the case of polling for synchronization).

\noindent \textbf{Synchronization.}
Most of the described security use cases unlock their true value if \acp{sedt} are synchronized with their corresponding \acp{cps}.
However, problems related to synchronization remain challenging for the community.
For example, defining an initial state for synchronization and identifying which system inputs serve as stimuli and, hence, need to be replicated in the virtual environment requires further investigation. 
Other important aspects to consider are real-time constraints and timing issues, which could lead to the state of the \ac{sedt} drifting from reality.
To handle such synchronization errors, mechanisms to recover from state mismatches are required.

\noindent \textbf{Security Implications of \acsp{sedt}.}
While the promises of \acp{sedt} may seem attractive, certain applications could also raise new concerns that potentially nullify the benefits.
For instance, in the case of intrusion detection, the \ac{sedt} may be susceptible to the same vulnerability as its physical counterpart.
Thus, attackers can potentially evade intrusion detection if malicious states are replicated, and the \ac{sedt} is equally affected by this vulnerability.
It should also not go unnoticed that \acp{dt}, by themselves, may introduce additional cybersecurity risks~\cite{Alcaraz2022}.
Consequently, careful consideration should be given to securing the \acp{dt} and their connection to their physical counterparts.

\section{Conclusion}
Overall, we conclude that the \ac{dt} paradigm holds promise for several \ac{cps} security applications.
However, the limitations discussed above suggest that considerable research and engineering efforts will be required.
Thus, \acp{sedt} will not render traditional security concepts and approaches obsolete in the near future, if ever.
Instead, we anticipate that they co-exist or that future \ac{sedt} technology will be used to further improve existing security solutions (e.g., testbeds, honeypots, \acp{ids}).
It is also worth highlighting that an \ac{sedt} is a software-only construct, meaning that its scope of application is naturally limited and, for instance, cannot serve as a substitute for HIL testing.
On a final note, we urge the scientific community to clearly communicate the characteristics of \acp{sedt} when presenting new solution proposals in order to prevent the nebulous view of the \ac{dt} concept from being prolonged.
\section*{Acknowledgment}
Matthias Eckhart was supported by
\begin{enumerate*}[label=(\roman*)]
    \item SBA Research (SBA-K1), which is a COMET Center within the COMET --- Competence Centers for Excellent Technologies Programme and funded by BMK, BMAW, and the federal state of Vienna. The COMET Programme is managed by FFG;
    \item SecurityTwin funded by the FFG via the BRIDGE 1 program under the grant number 880609.
\end{enumerate*}

Andreas Ekelhart was supported by CDL-SQI.
The financial support by the Christian Doppler Research Association, the Austrian Federal Ministry for Digital and Economic Affairs and the National Foundation for Research, Technology and Development is gratefully acknowledged.

Simin Nadjm-Tehrani and Magnus Almgren were supported by the RICS Centre on Resilient Information and Control Systems financed by Swedish Civil Contingencies Agency (MSB).

Mark Yampolskiy was supported by the U.S.~Department of Commerce, National Institute of Standards and Technology under Grants NIST-70NANB21H121 and NIST-70NANB19H170.

Furthermore, this work has benefited substantially from Dagstuhl Seminar 22171 ``Digital Twins for Cyber-Physical Systems Security.''

\bibliographystyle{IEEEtran}
\bibliography{IEEEabrv,bibliography}

\begin{IEEEbiography}{Matthias Eckhart}{\,}is a researcher with SBA Research, Vienna, Austria.
Contact him at meckhart@sba-research.org.
\end{IEEEbiography}

\begin{IEEEbiography}{Andreas Ekelhart}{\,}is a senior researcher with University of Vienna, Austria. 
Contact him at andreas.ekelhart@univie.ac.at.
\end{IEEEbiography}

\begin{IEEEbiography}{David Allison}{\,}is a researcher with AIT Austrian Institute of Technology, Vienna, Austria.
Contact him at david.allison@ait.ac.at.
\end{IEEEbiography}

\begin{IEEEbiography}{Magnus Almgren}{\,}is an associate professor with Chalmers University of Technology, Gothenburg, Sweden. 
Contact him at magnus.almgren@chalmers.se.
\end{IEEEbiography}

\begin{IEEEbiography}{Katharina Ceesay-Seitz}{\,}was a researcher with CERN, Geneva, Switzerland.
Contact her at katharina.ceesay-seitz@alumni.cern.
\end{IEEEbiography}

\begin{IEEEbiography}{Helge Janicke}{\,}is the research director of the Cyber Security Cooperative Research Centre, Joondalup, Australia and professor with Edith Cowan University, Joondalup, Australia.
Contact him at helge.janicke@cybersecuritycrc.org.au.
\end{IEEEbiography}

\begin{IEEEbiography}{Simin Nadjm-Tehrani}{\,}is professor with Linköping University, Sweden.
Contact her at simin.nadjm-tehrani@liu.se.
\end{IEEEbiography}

\begin{IEEEbiography}{Awais Rashid}{\,}is professor with University of Bristol, United Kingdom.
Contact him at awais.rashid@bristol.ac.uk.
\end{IEEEbiography}

\begin{IEEEbiography}{Mark Yampolskiy}{\,}is associate professor with Auburn University, Alabama, United States.
Contact him at mark.yampolskiy@auburn.edu.
\end{IEEEbiography}

\end{document}